# Transient-grating single-shot supercontinuum spectral interferometry (TG-SSSI)

S. W. Hancock, S. Zahedpour, H. M. Milchberg

*Institute for Research in Electronics and Applied Physics, University of Maryland, College Park, Maryland, 20742, USA*

*Abstract:* We present a technique for the single-shot measurement of the space- and time-resolved spatiotemporal amplitude and phase of an ultrashort laser pulse. The method, transient-grating single-shot supercontinuum spectral interferometry (TG-SSSI), is demonstrated by the space-time imaging of short pulses carrying spatiotemporal optical vortices (STOVs). TG-SSSI is well-suited for characterizing ultrashort laser pulses that contain singularities associated with spin/orbital angular momentum or polarization.

## I. Introduction

The need to characterize ultrashort laser pulses has spawned a large and increasing number of single-shot techniques including autocorrelation [1], multiple versions of frequency resolved optical gating (FROG) [2-6], spectral phase interferometry for direct electric-field reconstruction (SPIDER) and related methods [7-12], STRIPED FISH [13], d-scan [14], plus single-shot supercontinuum spectral interferometry (SSSI) [15-17] and other spectral interferometry methods [18,19]. While the basic FROG and SPIDER techniques extract only the space-independent temporal amplitude and phase, more complicated techniques [12-14] have recovered the *spatiotemporal* phase and amplitude of a laser pulse in a single-shot, albeit only with simple features such as pulsefront tilt. STRIPED-FISH [13] and d-scan [14] methods use iterative algorithms which, to our knowledge, have not been shown to converge for complicated structured light containing singularities, and SEA-SPIDER requires ancillary assumptions in determining the timing of spatial slices [12]. While SSSI does not recover the spatiotemporal phase of a pulse, it does recover the spatiotemporal pulse envelope, which has enabled measurement of ionization rates and ultrafast plasma evolution [20], electronic, vibrational and rotational nonlinearities [21,22], as well as nonlinear refractive indices and pulse front tilt [23].

In this paper, we present a new method that can measure, in a single-shot, the spatiotemporal phase and amplitude of an ultrafast laser pulse. It was developed for recent measurements [24] of pulses embedded with spatiotemporal optical vortices (STOVs) [25] and is well-suited for characterizing ultrashort laser pulses that contain singularities associated with spin/orbital angular momentum (SAM/OAM) [26-28] or polarization [29].

## II. Experimental Setup

We first briefly review SSSI by examining three of the beams in Fig. 1: the "structured pulse" $E_S$ which we want to measure, the reference pulse $E_{ref}$, and the probe pulse $E_{pr}$. Here, the structured pulse has spatiotemporal phase and amplitude imposed by the zero dispersion $4f$ pulse shaper [30-32] in the lower left of the figure. The reference and probe supercontinuum (SC) pulses are generated upstream of Fig. 1 by filamentation in a 2 atm $SF_6$ cell followed by a Michelson interferometer (not shown), with $E_{ref}$ leading $E_{pr}$ by ~ 2 ps. The transient amplitude of $E_S$ is measured via the phase modulation it induces in a spatially and temporally overlapped SC probe pulse $E_{pr}$ in an instantaneous Kerr "witness plate", here a thin (100-500 μm) fused silica window. The resulting spatio-spectral phase shift $\Delta\varphi(x,\omega)$ imposed on the probe is extracted from interfering $E_{pr}^{out} \sim \chi^{(3)} E_S E_S^* E_{pr}^{in}$ with $E_{ref}$ in an imaging spectrometer. Here, $E_{pr}^{in}$ and $E_{pr}^{out}$ are the probe fields entering and exiting the witness plate, $\chi^{(3)}$ is the fused silica nonlinear susceptibility, and $x$ is position within a 1D transverse spatial slice through the pump pulse at the witness plate (axes shown in Fig. 1). Fourier analysis of the extracted $\Delta\varphi(x,\omega)$ [20] then determines the spatio-temporal phase shift $\Delta\phi(x,\tau) \propto |E_S(x,\tau)|^2 \propto I_S(x,\tau)$, yielding the 1D space + time spatio-temporal intensity envelope $I_S$.

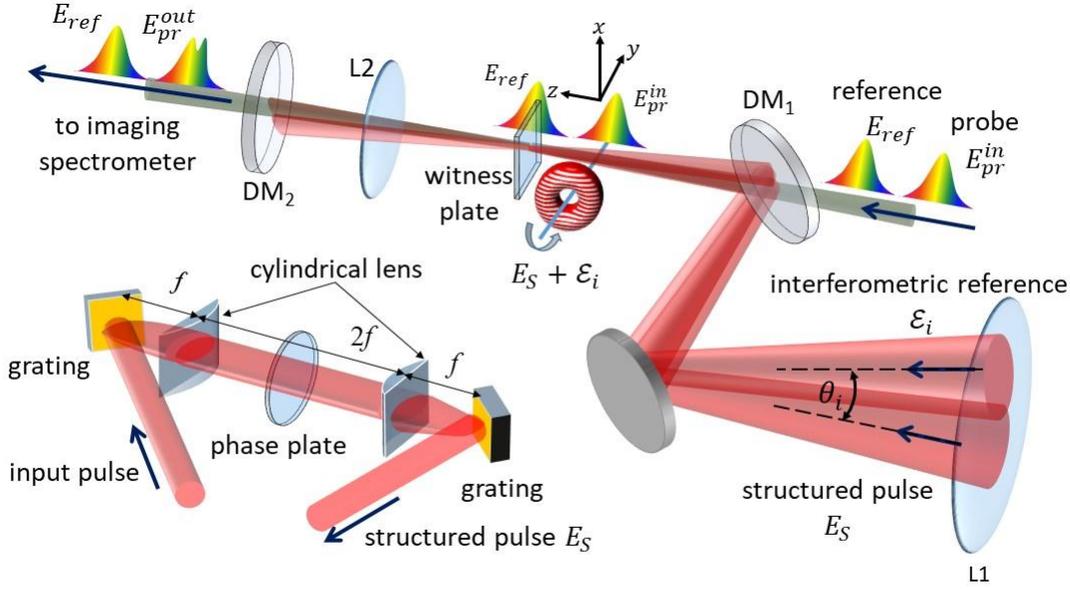

**Fig. 1.** Setup for transient grating single-shot spectral interferometry (TG-SSSI). The structured pulse $E_S$ and the interferometric reference pulse $\mathcal{E}_i$ cross at angle $\theta_w$ at the focus of lens L1 in a fused silica witness plate ($\theta_i$ in air), where a transient grating is formed. The grating is probed by a supercontinuum (SC) probe $E_{pr}^{in}$, preceded ~2ps earlier by a reference SC pulse $E_{ref}$. Imaged by L2, $E_{ref}$ and $E_{pr}^{out}$ interfere at an imaging spectrometer and the interferogram is analyzed in the Fourier domain, yielding a single-shot time and space resolved image of amplitude and phase of $E_S$. *Bottom left*: $4f$ pulse shaper [30-32] for generating spatiotemporally structured pulses $E_S$, here STOVs [24,33,34] imposed on a 50 fs, λ=800nm input pulse. The STOV phase windings are imposed by $l = \pm 1$, and $l = -2$ spiral phase plates in the Fourier plane of the pulse shaper. The phase plates are etched on fused silica and have 16 levels (steps) every $2\pi$.

Measurement of the spatiotemporal phase of $E_S$ is enabled by the addition of an interferometric reference pulse $\mathcal{E}_i$, which is crossed with $E_S$ at a small angle $\theta_i$ ($\theta_w = 3.15°$ in the witness plate). This forms a nonlinear transient refractive index grating, where $\mathcal{E}_i$ has the same center wavelength as $E_S$ but is bandpassed to be temporally longer. The transient grating (TG) is now the signal probed by SSSI (yielding the new method we call TG-SSSI), where the output probe pulse from the witness plate becomes $E_{pr}^{out} \propto \chi^{(3)} E_S \mathcal{E}_i^* E_{pr}^{in}$. The interference of $E_{pr}^{out}$ and $E_{ref}$ in the imaging spectrometer then enables extraction of $\Delta\varphi(x,\omega)$, yielding $\Delta\phi(x,\tau)$ as before. We note that $\Delta\phi(x,\tau)$ is the envelope of $E_S$ modulated by the transient grating: $\Delta\phi(x,\tau) \propto I_S(x,\tau) f(x,\tau)$, where $f(x,\tau) = \cos(2k_w x \sin(\theta_w/2) + \Delta\Phi(x,\tau))$ is the transient grating, $k_w = n_0 k$ is the pump central wavenumber in the witness plate, $n_0$=1.45, and $\Delta\Phi(x,\tau)$ is the spatiotemporal phase of $E_S$.

In the analysis of the 2D $\Delta\phi(x,\tau)$ images, $\Delta\Phi(x,\tau)$ is extracted using standard interferogram analysis techniques [13,14], and $I_S(x,\tau)$ is extracted using a low pass image filter (suppressing the sidebands imposed by the transient grating). Due to group velocity mismatch in the witness plate between $E_S$ (centre wavelength $\lambda_0 = 800$ nm) and the SC probe ($\lambda_{pr} = 600$nm), the extracted phase shift is smeared slightly in time by ~4 fs per 100µm of fused silica.

The laser used in the experiments is a 4 mJ/pulse, 50 fs FWHM, $\lambda_0 = 800$ nm, 1 kHz Ti:Sapphire system. The beam is split 3 ways, with ~100 µJ directed to SC generation (400-700nm) for $E_{pr}$ and $E_{ref}$, and a portion of the rest for $E_S$ and $\mathcal{E}_i$, whose energies were controlled using $\lambda/2$ plates and thin-film polarizers. The structured pulse $E_S$ was embedded with spatiotemporal phase windings by placing $l = \pm 1$ or $l = -2$ spiral phase plates at the Fourier plan of the $4f$ pulse shaper [24].

As depicted in Fig. 1, the SC reference and probe pulses, $E_{ref}$ and $E_{pr}^{in}$, are combined collinearly with the pulse $E_S$ using dichroic mirror DM$_1$, with $E_S$, $\mathcal{E}_i$, and $E_{pr}^{in}$ overlapping temporally in the witness plate, while $E_{ref}$ precedes them (by 2 ps). From the output face of the witness plate, $E_{ref}$ and $E_{pr}^{out}$ were magnified and relay imaged onto the spectrometer slit using high numerical aperture (NA) telescope with achromatic lenses. The large NA is necessary to collect the first order diffraction ($m = \pm 1$) of $E_{pr}^{out}$ from the transient grating. It is important for the imaging lenses to be achromatic for the image at the spectrometer slit to be in focus for all SC wavelengths and to minimize spherical aberration, which could spatially offset the diffracted orders of $E_{pr}^{out}$ from the zero order. Background and signal data were collected at 40 Hz by placing a chopper in the path of $E_S$, which allowed for the subtraction of the phase shift induced by $\mathcal{E}_i$ in the witness plate.

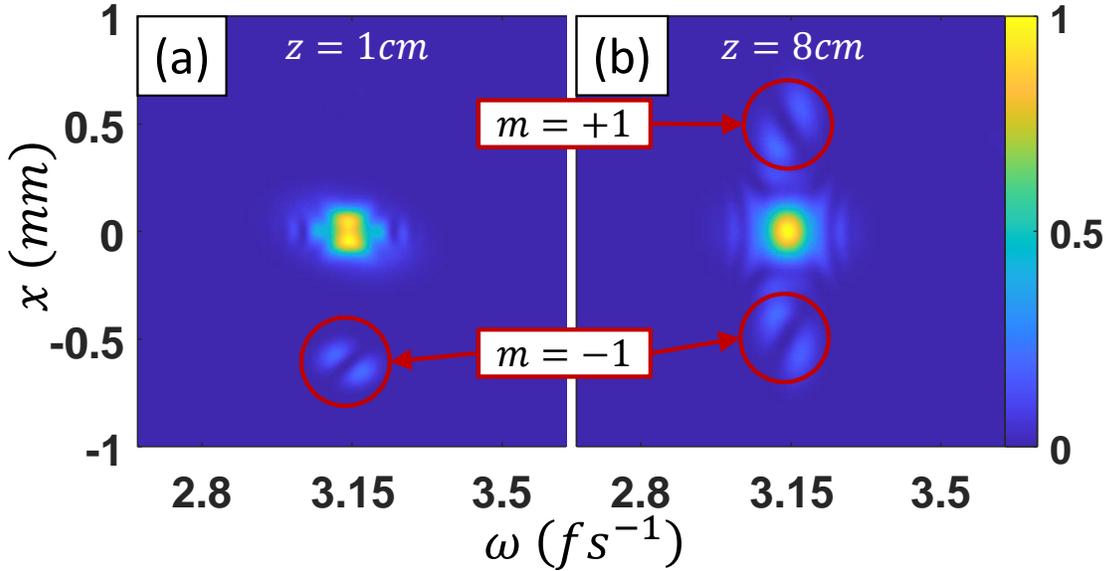

**Fig. 2.** Simulated spectrally-resolved scattering of supercontinuum probe pulse $E_{pr}$ from a transient nonlinear grating in a 500 µm thick fused silica plate, generated by the interference of pulses $E_S$ and $\mathcal{E}_i$, where $z$ is distance from the output face of the plate. Plotted as $|\Delta E_{pr}|^2 = |E_{pr}^{out} - E_{pr}^{in}|^2$. (a) Bragg regime transient grating: $\theta_w = 3.15°$, grating period $\Lambda = 10\mu m$, and $Q = 13.0$. Here, the scattering is captured at $z = 1$ cm owing to the rapid escape of the single diffracted order ($m = -1$) from the simulation window. (b) Raman-Nath regime transient grating: $\theta_w = 0.31°$, $\Lambda = 100\mu m$, and $Q = 0.13$, showing $m = \pm 1$ diffraction orders.

In principle, achromatic imaging of all diffracted orders precludes the need for detailed analysis of the diffraction. However, it is interesting to note that in our experiment, we observe only the zero order ($m = 0$) and the $m = -1$ order diffraction of the probe. To understand this, we assess whether probe diffraction is in the Bragg regime (one dominant diffraction peak) or in the Raman-Nath regime (multiple positive and negative diffraction orders) [35] by considering the dimensionless parameter $Q = 2\pi\lambda_i L/\Lambda^2 \bar{n}$, where $\lambda_i$ is the vacuum wavelength of incident light, $\Lambda$ is the interference grating period, $\bar{n}$ is the mean refractive index, and $L$ is the grating thickness. From ref. [35], diffraction is in the Raman-Nath regime for $Q \leq 1$ and in the Bragg regime for $Q \gg 1$. Our TG-SSSI configuration (with a $l = 500$ μm fused silica witness plate, $\Lambda(\theta_w = 3.15°) = 10$ μm, $n_2 = 2.5 \times 10^{-16}$ cm$^2$/W, and $\bar{n} = n_0 + n_0 n_2 I = n_0 + \Delta\phi_{TG}/kL$, where $\Delta\phi_{TG}$ is the modulated phase shift amplitude of the transient grating) gives $Q \cong 13.0$, which is in the Bragg regime. (Both $\Delta\phi_{TG}$ and $\Lambda$ are from our measurements.) This explains the observation of only one diffracted order.

This result is confirmed by simulations of scattering of $E_{pr}$ from the nonlinear grating formed by the interference of $E_S$ and $\mathcal{E}_i$. The simulation uses our implementation [36] of the unidirectional pulse propagation equation (UPPE) method [37], where all 3 beams intersect in the 500μm thick fused silica plate (with $E_S$ and $\mathcal{E}_i$ crossing at angle $\theta_w$ and $E_{pr}^{in}$ normal to the surface). The beam parameters are $E_S$ (λ=800nm, 50 fs FWHM, $w_0 = 100 \mu m$, $I_{S,peak} = 28$ GW/cm$^2$, $l = +1$ STOV), $\mathcal{E}_i$ (λ=800nm, 300 fs FWHM, $w_0 = 300 \mu m$, $I_{i,peak} = 28$ GW/cm$^2$, GDD = 0 fs$^2$), and $E_{pr}^{in}$ ($\lambda = 600$nm, $\Delta\lambda = 350$nm, 2.4 ps FWHM, $w_0 = 500 \mu m$, GDD = 1200 fs$^2$, TOD = 200 fs$^3$). The output electric field is numerically propagated 4 cm beyond the witness plate in air and then $E_S$ and $\mathcal{E}_i$ are spectrally filtered out, leaving the field $E_{pr}^{out}$. Figure 2(a) shows simulation results of probe diffraction for conditions similar to our experimental parameters ($\theta_w = 3.15°$ and $Q = 13.0$), where only the $m = -1$ diffraction order is present (Bragg regime), agreeing with our experiments. The crossing angle for Fig. 2(b) was chosen to be $\theta_w = 0.31°$, giving $Q = 0.13$, in the Raman-Nath regime, and the $m = \pm 1$ orders are present. We note that our current TG-SSSI setup could be adjusted to operate in the Raman-Nath regime by increasing the grating period $\Lambda$ (to $\Lambda \geq \sqrt{2\pi\lambda_i L/\bar{n}}$), but increasing the intensity of $E_S$ or $\mathcal{E}_i$ to increase $n_1$ could result in non-negligible plasma formation in the witness plate and refractive distortion of $E_{pr}$.

Figure 3(a) shows an example of a raw TG-SSSI interferogram frame recorded on the imaging spectrometer camera. Here, the pulse shaper generates $E_S$ as a $l = -2$ spatiotemporal optical vortex (STOV) pulse [24,25]. The vertical spectral fringe spacing is set by the Michelson-imposed time delay between the $E_{ref}$ and $E_{pr}^{out}$ pulses. The 2D phase shift $\Delta\varphi(x,\omega)$ is extracted in the same way as with all other SSSI interferograms [14,15], yielding $\Delta\phi(x,\tau) \propto I_S(x,\tau) f(x,\tau)$, which is plotted in Fig. 3(b). Here, the horizontal fringes imposed by $f(x,\tau)$ show the time-dependent interference between $E_S$ and $\mathcal{E}_i$. The spatiotemporal pulse envelope is recovered by low pass image filtering of $\Delta\phi(x,\tau)$ to remove $f(x,\tau)$, yielding $I_S(x,\tau)$ in Fig. 3(c).

Extraction of the spatiotemporal phase $\Delta\Phi(x,\tau)$ is performed by Fourier analysis along $x$ [38],

$$\Delta\Phi(x,\tau) = \arg(\mathcal{F}_k^{-1}\{\mathcal{F}_x\{\Delta\phi(x,\tau)\}\Theta(k)\}), \tag{1}$$

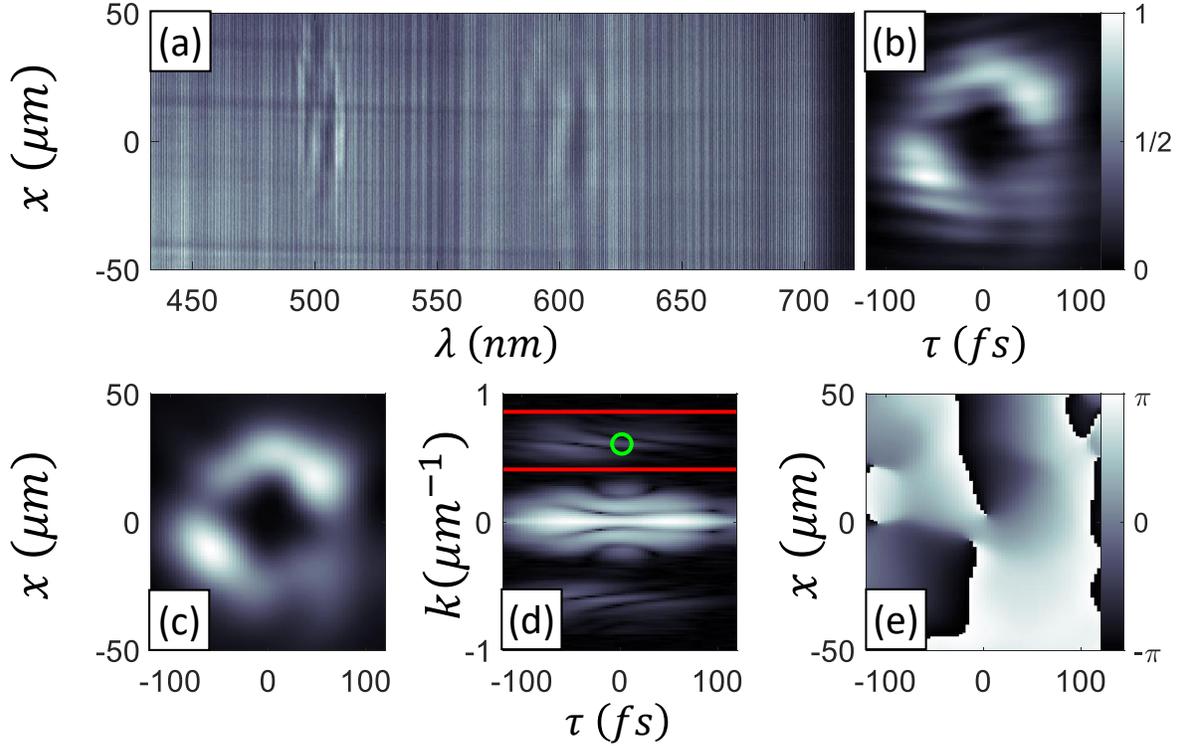

**Fig. 3.** Measurement of a $l = -2$ STOV-carrying pulse (interferometric reference $\mathcal{E}_i$ at 800nm and 10nm FWHM bandwidth); **(a)** Raw 1D space-resolved spectral interferogram; **(b)** extracted $\Delta\phi(x,\tau)$; **(c)** pulse envelope $I_S(x,\tau)$ from low pass filtering of $\Delta\phi(x,\tau)$; **(d)** $\log(|\mathcal{F}_x\{\Delta\phi(x,\tau)\}|+1)$. The red lines show the region to be spectrally windowed and the green circle identifies $(k_c, \tau_c)$ for frame averaging; **(e)** extracted spatiotemporal phase of the pulse, $\Delta\Phi(x,\tau)$.

where $\mathcal{F}_x\{\Delta\phi(x,\tau)\} = \Delta\tilde{\phi}(k,\tau)$ is the Fourier transform along $x$, $\mathcal{F}_k^{-1}\{\cdot\}$ is the inverse Fourier transform along $k$, $\Theta(k)$ is a sideband windowing and shifting ($k \to k - 2\pi/\Lambda$) function, and $k$ is the $x$-component of the spatial frequency. This is shown in Fig. 3(d) and (e). If the sideband is too close to the $k$-spectrum of the pulse envelope (which is centered at $k=0$), $\Theta(k)$ cannot separate the transient grating from the pulse envelope. This necessitates a larger spatial sample and/or finer grating period, considerations that have informed our pump-probe beam geometry.

Even though TG-SSSI is a single-shot method, averaging many shots of a reproducible transient process enables significant enhancement of the signal-to-noise ratio. Before averaging, however, the shot-to-shot shifting of the spatial interference fringes (from mechanical vibrations in the optical setup) must be compensated. The fringes are effectively forced into common alignment by adding a constant phase $\Delta\tilde{\phi}_n(k_c,\tau_c)$ to each frame, giving

$$\Delta\bar{\Phi}(x,\tau) = \arg\left(\frac{1}{N}\sum_{n=1}^{N}\mathcal{F}_k^{-1}\{[\Delta\tilde{\phi}_n(k,\tau) \times \exp(i\arg(\Delta\tilde{\phi}_n(k_c,\tau_c)))]\Theta(k)\}\right), \qquad (2)$$

where $\Delta\tilde{\phi}_n(k,\tau) = \mathcal{F}_x\{\Delta\phi_n(x,\tau)\}$, $\Delta\tilde{\phi}_n(k_c,\tau_c)$ is the constant phase added to frame $n$ to align the fringes, $(k_c,\tau_c)$ is a common point across all $N$ frames, and $\Delta\bar{\Phi}(x,\tau)$ is the mean spatiotemporal phase. The point $(k_c,\tau_c)$ is chosen at a location in the sideband where the signal is sufficiently larger than the phase noise, otherwise each frame would be offset by a random phase factor.

## III. Results and Discussion

To demonstrate TG-SSSI, we used the $4f$ pulse-shaper to generate (a) a Gaussian pulse ($l = 0$, no phase plate), and STOV-carrying pulses with topological charge (b) $l = +1$, (c) $l = -1$, and (d) $l = -2$, using corresponding spiral phase plates in the Fourier plane of the shaper. The columns of Fig. 4 show $\Delta\phi(x,\tau)$, $I_s(x,\tau)$, $f(x,\tau)$, and $\Delta\Phi(x,\tau)$ for pulses carrying $l = 0, \pm 1$, and $-2$. For $l = 0$ (row (a)), we see a slight fringe curvature in the transient grating $f(x,\tau)$, indicating a dispersion mismatch between $E_S$ and $\mathcal{E}_i$. For the $l = \pm 1$ STOVs in rows (b) and (c), $f(x,\tau)$ clearly shows the transient fringe fusing or splitting identifying the opposite phase windings shown in the $\Delta\Phi(x,\tau)$ column. For $l = -2$ (row (d)), one fringe in $f(x,\tau)$ splits into three at the center of the pulse. Upon phase extraction, $\Delta\Phi(x,\tau)$ has two nearby $l = -1$ phase windings rather than a single $l = -2$ winding. We attribute this to a mismatch between the transverse beam dimensions at the Fourier plane of the pulse shaper and the radially independent phase winding of the $l = -2$ phase plate. Since the profile of the beam in the Fourier plane of the shaper (itself dictated by the grating periodicity and cylindrical lens focal length) is slightly elliptical, one of the axes of the phase plate should ideally be scaled to match the ellipticity of the beam. Utilizing a programmable spatial light modulator rather than a fixed phase plate in the pulse shaper would enable scaling of the phase mask to match the beam profile, making possible the generation of $l = \pm 2$ and even higher order STOVs.

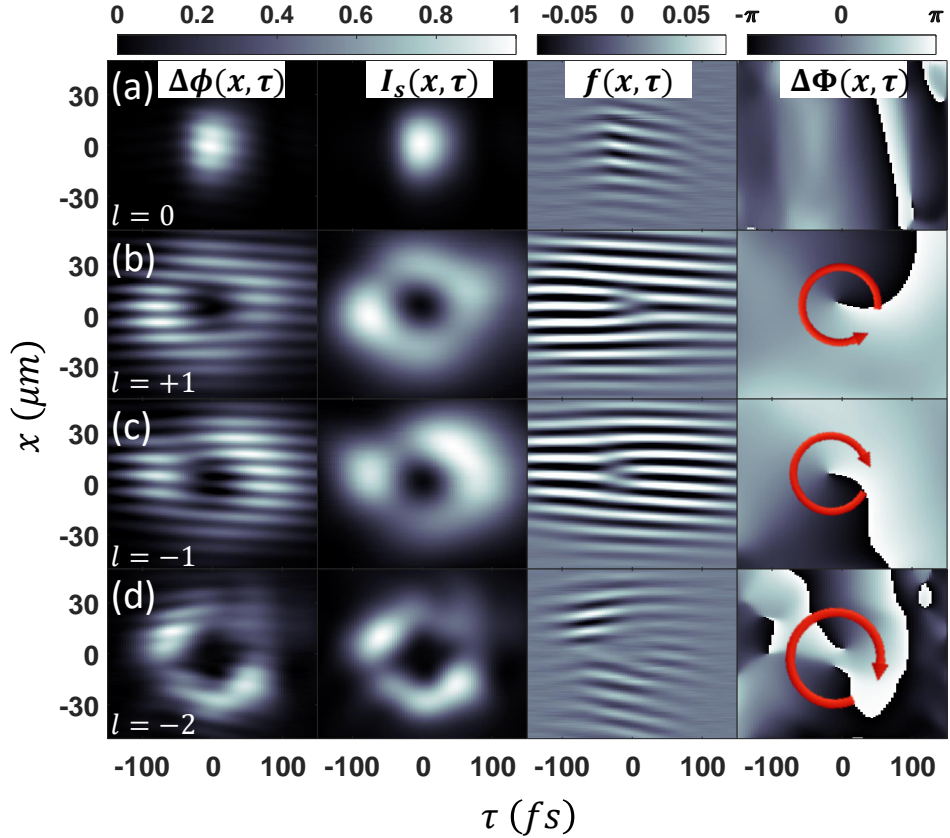

**Fig. 4.** Experimental results from TG-SSSI after fringe alignment. Columns show the extracted full TG-SSSI signal $\Delta\phi(x,\tau)$, which is low-pass filtered to yield the pulse intensity envelope $I_s(x,\tau)$, or high-pass filtered to give the transient grating $f(x,\tau)$, from which the spatiotemporal phase $\Delta\Phi(x,\tau)$ is extracted. The rows show results for (a) a Gaussian pulse ($l = 0$), (b) a $l = +1$ STOV, (b) a $l = -1$ STOV, and (c) results from an $l = -2$ phase plate. The red arrows denote the direction of increasing spatiotemporal phase.

## IV. Conclusion

In summary, we have presented a new single-shot diagnostic of ultrashort spatio-temporally structured laser pulses, transient grating single-shot supercontinuum spectral interferometry (TG-SSSI) and have used it to measure simple Gaussian and STOV-carrying pulses generated by a $4f$ pulse-shaper. Among multiple possible applications, TG-SSSI should prove useful in the study of nonlinear propagation, collapse and collapse arrest of intense laser pulses in transparent media, where spatio-temporal optical pulse structures naturally emerge [25]. Finally, we note for experiments where the structured pulse $E_S$ is highly repetitive and reproducible, TG-SSSI could be extended to two spatial dimensions ($x$ and $y$) by transversely scanning $E_{pr}^{out}$ across the spectrometer entrance slit in the $y$ direction, as is done in 2D+1 SSSI [20], to obtain $I_S(x, y, \tau)$ and $\Delta\Phi(x, y, \tau)$.

**Acknowledgements** The authors thank I. Larkin for discussions and technical assistance. This work is supported by Air Force Office of Scientific Research (FA9550-16-10121, FA9550-16-10284); Office of Naval Research (N00014-17-1-2705, N00014-20-1-2233); National Science Foundation (PHY2010511).